\documentclass[twocolumn, prb]{revtex4}

\usepackage{graphicx}
\usepackage{dcolumn}
\usepackage{bm}

\begin{document}

\title{ Quantum phase transition and sliding Luttinger liquid  
in coupled $t-J$ chains}

\author {S. Moukouri }

\affiliation{ Department of Physics and  Michigan Center for 
          Theoretical Physics \\
         University of Michigan 2477 Randall Laboratory, Ann Arbor MI 48109}

\begin{abstract}
 Using a recently proposed perturbative numerical renormalization-group
algorithm, we explore the connection between quantum criticality and the
emergence of Luttinger liquid physics in $t-J$ chains coupled by frustrated
interactions. This study is built on an earlier finding that at the maximally 
frustrated point, the ground state of weakly-coupled Heisenberg chains is 
disordered, the transverse exchanges being irrelevant. This result is extended 
here to transverse couplings up to $J_{\perp}=0.6$, and we argue that it may also
be valid at the isotropic point. A finite size analysis 
of coupled Heisenberg chains in the vicinity of the maximally frustrated point
 confirms that the transverse spin-spin correlations decay exponentially
while the longitudinal ones revert to those of decoupled chains. We find that 
this behavior  persists upon moderate hole doping $x \alt 0.75$. For larger 
doping, the frustration becomes inactive and the quantum critical point 
is suppressed. 
\end{abstract}

\maketitle

\section{Introduction}

 In a number of materials including cuprate high temperature 
superconductors \cite{reviewHTC},  organic \cite{reviewOC} and 
inorganic \cite{reviewIOC} quasi one-dimensional (1D) conductors, the 
physics is dominated by the combined effect of low dimensionality, strong
electron correlations and competing orders. These materials 
display a phenomenology, in the metallic state above the ordered states,
which departs from the Fermi liquid (FL) picture. This has motivated a search
for a new paradigm which goes beyond FL theory. Two new ideas, electron 
fractionalization \cite{anderson,kivelson0} and quantum 
criticality \cite{sachdev} (QC), have been proposed as the possible 
physical effects which may lead to the observed non-FL behavior.

The breakdown of the FL theory
is known to occur in 1D where the low energy physics is described by
Luttinger liquid (LL) theory. The LL is fundamentally different from a FL; 
for instance, instead of the quasiparticle peak displayed by the FL at zero 
energy in the spectral weight function, a pseudogap is
observed in the LL. This is because in the LL the low energy excitations  
are collective spin density and charge density, in contrast to the FL where
they are still electron-like. An interpretation of this phenomenon is that
the electron fragments into density waves known as spinons and holons which
propagate with different velocities. The LL would thus be the natural starting 
point to search for non-FL behavior in higher dimension. However, it has been 
impossible to find a LL in dimensions greater than one using perturbative 
methods. This has been tried by starting from 1D models  coupled  by a 
small transverse hopping \cite{bourbonnais} or  
directly from the 2D non-interacting electron gas in which the interaction
is introduced perturbatively\cite{shankar}. It was found in both cases that 
the FL is the stable fixed point. An alternative strategy applied recently 
is to build a LL fixed point in anisotropic 2D systems by coupling 1D LLs 
by marginal transverse strong forward density-density or current-current 
couplings \cite{kivelson,carpentier,kane}. 
The new 2D fixed point, called sliding LL (SLL), retains the physics of the 1D 
LL. It was shown that there is a regime of parameters in which Josephson, 
charge and spin density waves and single particle hopping are irrelevant, 
i.e., a domain of stability of the 2D LL. There has not been, to the author's 
knowledge, any detailed analysis as to why these two approaches lead to 
conflicting conclusions. It would appear that using simple perturbation theory,
 one cannot go smoothly from a single chain to a SLL.

A different approach has been to assign the departure from a FL behavior to 
the proximity of a quantum critical point (QCP)\cite{sachdev}. 
Because of the proximity of the QCP the adiabatic continuity concept, which 
is the basis of the FL theory, is no longer valid. In the normal state, 
the eigenstates of the system are a mixture of states which evolve 
separately into the ground states existing on the two sides of the QCP.
Because of this mixture, the single-particles excitations of the system 
are not simple electron-like excitations as in the FL.

Considering the pure spin one-half limit of the problem, various
studies have found that any small exchange leads to the onset of
long-range order. Laughlin \cite{laughlin} has argued that spin-$1/2$ 
systems have the propensity to order and it is only at a QCP that they 
do not. Yet considering that experiments clearly
show a departure from the FL, one can still retain the electron 
fractionalization hypothesis as  the driving mechanism of the non-FL but 
 attribute it not to LL physics but to the presence of a QCP. 
This suggests that in constructing a model Hamiltonian for coupled
LLs, one must include competing interactions that preclude any order
at the QCP. 
  
The difference between the SLL and QCP physics is ultimately
two different views of the same reality. If they are both  
correct, they should emerge from a well controlled study of an 
appropriate microscopic Hamiltonian. The two-step density-matrix 
renormalization group (TSDMRG) \cite{moukouri-TSDMRG, moukouri-TSDMRG2} 
offers a new approach to study these issues. Although it is a perturbative 
approach, it has the important non-perturbative property that it can lead 
to an ordered state starting from a disordered 
state\cite{moukouri-TSDMRG2, moukouri-KB}.
Thus it has the ability to treat effectively the different ground states that
may arise in weakly coupled chains.

In this paper, we apply the TSDMRG to
study the QPT in the spatially anisotropic $t-J-t_{\perp}-J_{\perp}-J_d$ model. 
 We show that in this model the concept of LL can be unified with that of
quantum criticality, namely that the model undergoes a QPT 
at $J_d = J_{\perp}/2$ (These couplings are defined in the Hamiltonian
(~\ref{hamiltonian}) given below). At the QCP the system is in a state made of 
nearly disconnected chains that can be interpreted as a SLL. We first analyze 
the half-filling case which corresponds to the anisotropic $J_1-J_2$ model.
We show that the dimerization found in previous approximate field theory
analyses \cite{tsvelik,starykh} does not occur.
We then study the doped case at  hole densities up to
 $x=0.4$. The picture observed in the half-filled case remains valid
at moderate dopings. At the QCP, the transverse
spin-spin correlation function decays exponentially as in the half-filled
case. But at higher dopings, the frustration becomes ineffective and 
the transition is suppressed; the ground state is a spin density wave at
all couplings. This suggests the existence of a SLL to FL crossover at finite 
temperature.

\section{Model and Method}

The anisotropic $t-J-t_{\perp}-J_{\perp}-J_d$ model is:
\begin{eqnarray}
 \nonumber H=-t\sum_{i,l}(c_{i,l}^{\dagger}c_{i+1,l}+h.c.)+ 
J \sum_{i,l}{\bf S}_{i,l}{\bf S}_{i+1,l}-\frac{1}{4}n_{i,l}n_{i+1,l}\\
\nonumber -t_{\perp}\sum_{i,l}(c_{i,l}^{\dagger}c_{i,l+1}+h.c.)+ 
 J_{\perp} \sum_{i,l}{\bf S}_{i,l}{\bf S}_{i,l+1}-
 \frac{1}{4}n_{i,l}n_{i,l+1} \\
+J_d \sum_{i,l}{\bf S}_{i,l}{\bf S}_{i+1,l+1}+{\bf S}_{i+1,l}{\bf S}_{i,l+1}
\label{hamiltonian}
\end{eqnarray}

\noindent where $t$ and $J>0$ are the in-chain hopping and exchange parameters;
$t_{\perp}$ and $J_{\perp}>0$ are the transverse hopping and exchange parameters;
$J_d >0$ is the diagonal exchange parameter. 
We start by describing the extension to fermion systems of the numerical 
renormalization method, which has been applied so far only to spin systems 
\cite{moukouri-TSDMRG,moukouri-TSDMRG2}.
There is no significant difficulty in going from spins to fermions. One
should simply be careful to consistently apply fermion anticommutation rules.
It is crucial to choose an order for the sites in the 2D lattice and keep
it throughout the derivation of the algorithm. 
 The method is a special case of a more general matrix perturbation 
method based on the Kato-Bloch expansion \cite{kato,bloch} which was recently
introduced by the author \cite{moukouri-KB}. The method has two main steps.
In the first step, the usual 1D DMRG method \cite{white} is applied
to find a set of low
lying eigenvalues $\epsilon_n$ and eigenfunctions $|\phi_n \rangle$ of a
single chain. In the second step, the  2D Hamiltonian is then projected
onto the basis constructed from the tensor product of the $|\phi_n \rangle$'s.
This projection yields an effective one-dimensional Hamiltonian for 
the 2D lattice,

\begin{eqnarray}
 \nonumber \tilde{H} \approx \sum_{[n]} E_{\parallel [n]} |\Phi_{\parallel [n]}
\rangle \langle\Phi_{\parallel [n]}| -
 t_{\perp}\sum_{i,l}(\tilde{c}_{i,l}^{\dagger}\tilde{c}_{i,l+1}+h.c.)+ \\
\nonumber J_{ \perp} \sum_{il} {\bf \tilde{S}}_{i,l} {\bf \tilde{S}}_{i,l+1}
-\frac{1}{4}\tilde{n}_{i,l}\tilde{n}_{i+1,l} \\ 
 J_{d} \sum_{il} {\bf \tilde{S}}_{i,l} {\bf \tilde{S}}_{i+1,l+1}+
{\bf \tilde{S}}_{i+1,l} {\bf \tilde{S}}_{i,l+1},
\end{eqnarray}

\noindent where  $E_{\parallel [n]}$ is the sum of eigenvalues of the
different chains, $E_{\parallel[n]}=\sum_l{\epsilon_{n_l}}$;
$|\Phi_{\parallel [n]}\rangle$ are the corresponding eigenstates,
$|\Phi_{\parallel [n]}\rangle =  |\phi_{n_1}\rangle  |\phi_{n_2}\rangle ...
|\phi_{n_L} \rangle$; $\tilde{c}_{i,l}^{\dagger}$, $\tilde{c}_{i,l}$, and
${\bf \tilde{S}}_{i,l}$ are the  renormalized matrix elements in the single 
chain basis. They are given by

\begin{eqnarray}
(\tilde{c}_{i,l}^{\dagger})^{n_l,m_l}=(-1)^{n_i}\langle \phi_{n_l}|{c}_{i,l}^{\dagger}
|\phi_{m_l}\rangle, \\
(\tilde{c}_{i,l})^{n_l,m_l}=(-1)^{n_i}\langle \phi_{n_l}|{c}_{i,l} |\phi_{m_l}\rangle, \\
{\bf \tilde{S}}_{i,l}^{n_l,m_l}=\langle \phi_{n_l}|
{\bf S}_{i,l}|\phi_{m_l}\rangle,
\end{eqnarray}

\noindent where $n_i$ represents the total number of fermions from sites $1$ to
$i-1$. For each chain, operators for all the sites are stored in a 
single matrix 

\begin{eqnarray}
\label{bm1} 
\tilde{c}_{l}^{\dagger}=(\tilde{c}_{1,l}^{\dagger},...,
\tilde{c}_{L,l}^{\dagger}),\\ 
\label{bm2} 
\tilde{c}_{l}=(\tilde{c}_{1,l},...,\tilde{c}_{L,l}),\\ 
{\bf \tilde{S}}_{l}=({\bf \tilde{S}}_{1,l},...,{\bf \tilde{S}}_{L,l}).
\label{bm3}
\end{eqnarray}

\noindent Since the in-chain degrees of freedom have been integrated out,
the interchain couplings are between the block matrix operators in 
Eq.(~\ref{bm1},~\ref{bm2},~\ref{bm3}) which depend only on the chain index $l$. 
In this matrix notation, the effective Hamiltonian is 
one-dimensional and it is also studied by the DMRG method. The only 
difference with a normal 1D situation is that the local operators are now 
$m \times m$ matrices, where $m$ is the number of states kept to describe 
the single chain. In this study, mostly exact diagonalization (ED) instead 
of DMRG is used in the 1D part of the algorithm. This has the obvious 
advantage that all the eigenvectors and eigenvalues are known up to a well 
defined accuracy which is set to $10^{-6}$ in the Davidson algorithm used to 
obtain them. The two-step DMRG method is variational 
\cite{moukouri-TSDMRG2, alvarez}, so the results can systematically be 
improved by increasing $m$. In this study, the calculations were done on a 
Dell 670 dual Xeon (3.4 GHz) workstation with 4GB of RAM. The maximum $m$ 
we can reach for the calculations to run within a reasonable time (about three
 days) is about $m=96$ for a $L\times(L+1)$ lattice with $L=16$.

In this study, we are mostly concerned with the parameter regime where 
$J_d \approx J_{\perp}/2$. It is in this regime that the sign problem
in the QMC is the most severe. In Ref.\cite{alvarez}, we
have shown that the two-step DMRG is at its best in the highly frustrated
 regime where QMC simulations cannot be done. To illustrate this fact, 
we show in Table(~\ref{dmrgvsqmc}) the ground state energy ($E_G$) as a
function of $m$ for the unfrustrated case, with $J_{\perp}=0.2$ and $J_d=0$,
 and in the frustrated case at the maximally frustrated point ( this point is
defined below), with $J_{\perp}=0.2$ and $J_d=0.114$, for a $12 \times 13$ 
lattice. The extrapolated value of $E_G$ in the unfrustrated case is  in 
good agreement with the QMC. It can be seen that $E_G$ converges faster in the 
frustrated
case. The strategy in this study is to increase $J_{\perp}$ while the
ratio $J_d/J_{\perp}$ remains close to $0.5$. We do not expect a significant
decrease in accuracy for larger $J_{\perp}$ because as we will see below, 
the interchain correlations decay exponentially in this region. Thus despite 
its perturbative nature, the TSDMRG is able to treat model(~\ref{hamiltonian}) 
well beyond the weak coupling regime {\it as long as one stays in the vicinity
 of the QCP}. 

\begin{table}
\begin{ruledtabular}
\begin{tabular}{ccc}
 $m/J_d$ & $0.$ & $0.114$  \\
\hline
 $32$ & 0.43796 & 0.42898 \\
 $64$ & 0.43928 & 0.42917  \\
 $96$ & 0.43970 & 0.42920 \\
$\infty$& 0.44051 & 0.42921 \\
 QMC  & 0.44075 &  \\
\end{tabular}
\end{ruledtabular}
\caption{Ground-state energies as function of $m$ and $J_d$ for
$J_{\perp}=0.2$.}
\label{dmrgvsqmc}
\end{table}

The high accuracy enjoyed at half-filling is somewhat reduced in the doped
case. The reason behind this reduction is simply that the size of the 
total Hilbert space increases as one goes from 
two degrees of freedom per site to three. Since we keep roughly the
same number of states in the two cases, the accuracy will be smaller
for doped systems. This can be seen, for instance, when $m=96$ states
are kept in the second step. The energy width of the retained states
drops from $\Delta E \approx 1.6$ at half-filling to $\Delta E \approx 0.8$
when the doping is $x=0.125$ for a $L=16$ system. Another difficulty which 
arises upon doping is that one needs to target, during the first step of the 
algorithm, at least three charge sectors in order to keep matrix elements 
which involve interchain hopping.  A typical situation is $L=16$ and $x=0.125$, where the
corresponding ground state has $14$ electrons. For this charge sector,
five spin sectors with $S_Z=0,\pm 1,\pm 2$ were targeted. For
charge sectors with $13$ and $15$ electrons, four spin sectors with $S_Z=\pm 0.5, \pm 1.5$ were also targeted. Unlike the half-filled case, ED was not done
for all doped systems in the first step. For instance for $L=16$, we kept
$243$ states. This number of states corresponds to an exact 
diagonalization of 12 sites which means three DMRG iterations were necessary
to reach the desired size. Hence, the truncation error at the end of the first
step increases from zero at half-filling to about $1 \times 10^{-6}$ in the
doped case. 

Removing or adding one electron on a finite system can significantly affect
the density if the system size is not large enough. For instance, during the
study of a $L=16$ lattice with the hole density $x=0.125$, the nominal
ground state has 14 electrons, whereas the ground states with  13 and
15 electrons correspond to the hole densitites $0.1875$ and $0.0625$ 
respectively. Thus, adding or removing holes on small systems induces large
variations of the hole density, resulting in large differences between
the ground states of the different charge sectors. These variations are
the largest near half-filling for small values of $J$. As a consequence, the
electrons cannot jump between the chains and the charge degrees of freedom 
remain decoupled for small values of the interchain hopping.  
It is necessary to fine tune $J$ so that these energy differences are smaller
or of the same magnitude as $t_{\perp}$. For $t_{\perp}=0.2$, which will be
used for all doped systems,  $J$ was chosen
so that the difference $\delta_c$ between ground state energy of charge 
sectors differing by one electron is such that $\delta_c \alt 0.16$. 
Fixing $\delta_c$ amounts to fixing the chemical potential instead of
the coupling. The value of $J$ which satisfies this criterion depends on the 
hole density $x$. These values are listed in Table~\ref{mu} for $L=16$ systems.

\begin{table}
\begin{ruledtabular}
\begin{tabular}{cccc}
 $x$ & $0.125$ & $0.25$ & $0.375$ \\
\hline
 $J$ & 1.4 & 1.14 & 0.75 \\
 $E(N)-E(N+1)$ & 0.1012 & 0.0982 & 0.1355  \\
 $E(N)-E(N-1)$ & 0.1250 & 0.1658 & 0.1615 \\
\end{tabular}
\end{ruledtabular}
\caption{Hole densities, corresponding exchange couplings and
energy differences between charge sectors for $L=16$.}
\label{mu}
\end{table}

\section{Finite size analysis at the critical point at half-filling}

The half-filled case corresponds to weakly coupled Heisenberg chains that
we studied in Ref\cite{moukouri-TSDMRG2}. In that study it was found that
the model displays two phases: a N\'eel $Q=(\pi,\pi)$ when 
$J_d/J_{\perp} \alt 0.5$ and a N\'eel $Q=(\pi,0)$ when $J_d/J_{\perp} \agt 0.5$.
These two phases are separated by a critical point where the long-distance
behavior of the spin-spin correlations along the chains in the 2D system
is identical to those in the disconnected chains. In this study, we will
show that this behavior found for small $J_{\perp} \alt 0.2$ extends well
beyond the weak interchain coupling regime (up to $J_{\perp}=0.6$).
The nature of the ground state in this regime, which was previously 
called "nearly
independent chains", will be further clarified by the analysis of the 
interchain spin-spin correlations.   
 Before starting the finite size analysis, it is important to stress the
fact that the open boundary conditions (OBC) which were applied 
in most of this study create some difficulty
in the interpretation of data. While OBC have the advantage of yielding 
higher accuracy in DMRG simulations, they
artificially break the lattice translational symmetry. 
 The OBC thus introduce a spurious dimerization $d$ in the 1D chain, 
as we will see below, which converges very slowly toward zero. 
To our knowledge, the exact behavior of this dimerization as a function 
of the system size has not  been reported in the literature. On top of
this difficulty, one must account for logarithmic corrections
to the finite size quantities. For instance, fitting $d$ with  $1/L$, 
one finds that $d$ extrapolates to a finite value in the limit 
$L \rightarrow \infty$. This deviation from $0$, which is the true limit,
is the signature of logarithmic corrections. Logarithmic corrections in an
open Heisenberg chain have been studied in Ref.~\cite{affleck}. They are due to
marginally irrelevant operators and their mathematical form is hard to 
find exactly for small systems. 
A functional form exists only in the limit of large $L$. For instance, the
finite size spin gap has the following form

\begin{equation}
\Delta_L \approx \frac{\pi v}{L}[1-\frac{4\pi g(L)}{\sqrt{3}}]
\end{equation}

\noindent where $v=\pi/2$ is the spin velocity and $g(L)$ a function 
such that

\begin{equation}
g(L) \rightarrow \frac{\sqrt{3}}{4\pi \log{L}}
\end{equation} 

\noindent when $L$ is very large. For small chains, $g(L)$ has to be introduced
as an unknown parameter which is obtained by fitting the data to known
 limits. Since the main interest of this study lies 
 in the relative behavior of the 2D system with respect to an isolated
 1D chain, we 
will not use $g(L)$ during the finite size analysis. 
 When  performing the finite size analysis, simple linear or
quadratic functions of $1/L$ will be used even though they do not lead
to the exact values in the thermodynamic limit for the 1D system because of
the logarithmic corrections.  

\subsection{Ground-state energies}


\begin{figure}
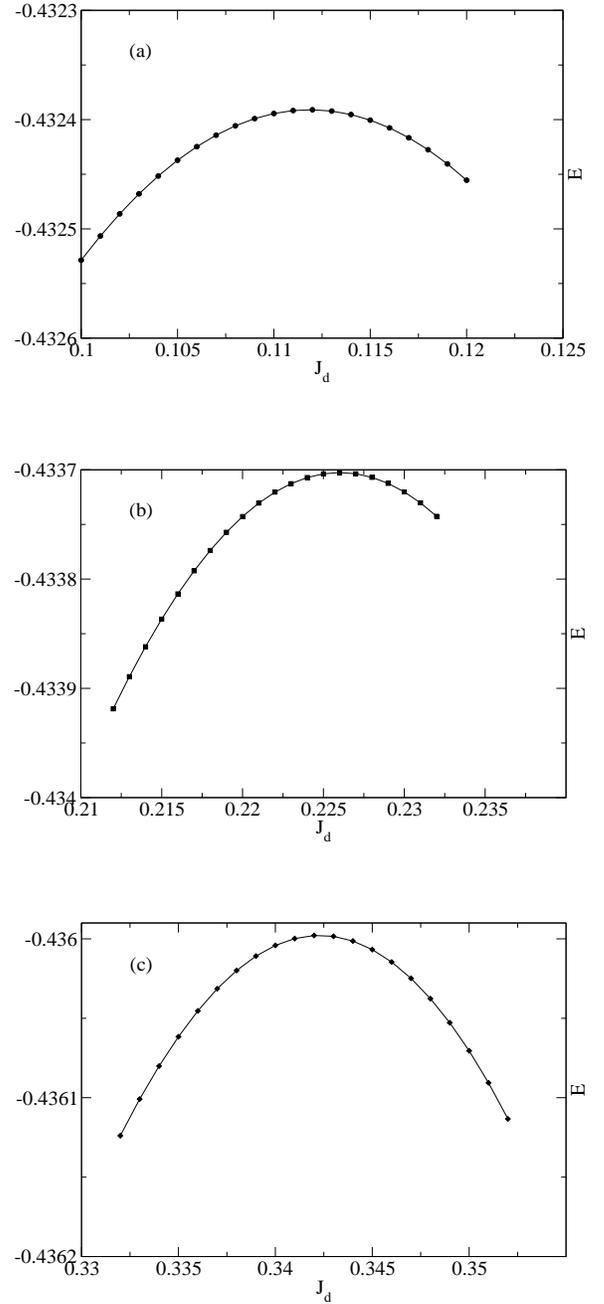

\includegraphics[width=3. in, height=2. in]{gsg0.2l16.eps}\\
\vspace{1cm}
\includegraphics[width=3. in, height=2. in]{gsg0.4l16.eps}\\
\vspace{1cm}
\includegraphics[width=3. in, height=2. in]{gsg0.6l16.eps}\\
\caption{Ground-state energies for $L=16$ and $J_{\perp}=0.2$(a),
$J_{\perp}=0.4$ (b) and, $J_{\perp}=0.6$ (c) as function of $J_d$.}
\vspace{0.5cm}
\label{gs}
\end{figure}

\begin{figure}
\includegraphics[width=3. in, height=2. in]{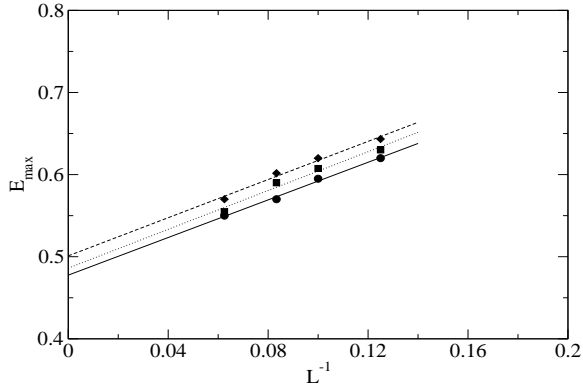}\\
\caption{ Maxima of the ground-state
energies as function of the lattice size for $J_{\perp}=0.2$ (circles),
$J_{\perp}=0.4$ (squares) and, $J_{\perp}=0.6$ (diamonds).}
\vspace{0.5cm}
\label{gsmax}
\end{figure}

\begin{table}
\begin{ruledtabular}
\begin{tabular}{ccccc}
 $J_{ \perp}$ & $L=8$ & $L=10$ & $L=12$ & $L=16$  \\
\hline
 $0.2$ & 0.124 & 0.119 & 0.114 & 0.112 \\
 $0.4$ & 0.252 & 0.243 & 0.236 & 0.226 \\
 $0.6$ & 0.386 & 0.372 & 0.361 & 0.342 \\
\end{tabular}
\end{ruledtabular}
\caption{Values of $J_d$ corresponding to the maximum of the
ground-state energy for different $J_{\perp}$ and $L$. }
\label{maxgs}
\end{table}

The ground-state energies shown in Fig.(~\ref{gs}) show similar behavior
for all values of $J_{\perp}$ when $J_d$ is varied. Starting from
$J_d=0$ where the ground state is a N\'eel $Q=(\pi,\pi)$ phase,
$E_G$ typically increases with $J_d$ until it reaches a maximum, 
labeled the maximally frustrated point $J_d^{max}$. Then $E_G$ decreases
when $J_d > J_d^{max}$ where the ground state is in a N\'eel 
$Q=(\pi,0)$ phase. The ratio $J_d^{max}/J_{\perp}$
depends on both $J_{\perp}$ and $L$. For small $J_{\perp}$, 
$J_d^{max}/J_{\perp}$ remains very close to $0.5$ but deviates from this
value as $J_{\perp}$ increases. Previous studies ( see Ref.\cite{lhuillier}
for a review) found
$J_d^{max}/J_{\perp}=0.6$ when $J_{\perp}=1$. This deviation is larger
for OBC than for PBC. This is because the maximally frustrated point
corresponds to the point where the transverse component of the Fourier 
transform of the coupling

\begin{equation}
J(Q)= 2\cos{Q_x}+ 2\cos{Q_y}[J_{\perp}+2J_d \cos{Q_x}]
\end{equation}

\noindent vanishes when $Q_x=\pi$. At this point $2 J_d$ bonds
cancel one $J_{\perp}$ bond. But when OBC are used on a 
$L\times(L+1)$ lattice, one has $L^2$ bonds but only $2L(L-1)$ 
$J_d$ bonds. Thus for small lattices, there is a deficit in $J_d$
bonds which translates into the shift of the maximum towards larger
$J_d$ for a fixed $J_{\perp}$.

This shift has been argued to be indicative of the existence of an 
intermediate phase, possibly dimerized, in the region between 
$J_d=0.4$ (where the N\'eel
$Q=(\pi,\pi)$ order parameter was found to vanish) and $J_d=0.6$ (where
the N\'eel order parameter $Q=(\pi,0)$ or $(0,\pi)$ vanishes). Our results
shown in Table(~\ref{maxgs}), which list the position of the maxima
for various $J_{\perp}$ and $L$, indicate that this deviation of 
$J_d^{max}/J_{\perp}$ from $0.5$ is simply a finite size effect. 
For a fixed $L$, $J_d^{max}/J_{\perp}$ increases with $J_{\perp}$ in
agreement with the finding in the isotropic case. But one can see that
for a fixed $J_{\perp}$, $J_d^{max}/J_{\perp}$ decreases with $L$.
Fig.(~\ref{gsmax}) displays the extrapolated values of $J_d^{max}/J_{\perp}$
for $J_{\perp}=0.2$, $0.4$, and $0.6$. They all converge to the vicinity
of $0.5$. Aside from numerical uncertainties, it is thus likely that 
$J_d^{max}/J_{\perp}=0.5$ for all $J_{\perp}$  in the thermodynamic limit.

\begin{figure}
\includegraphics[width=3. in, height=2. in]{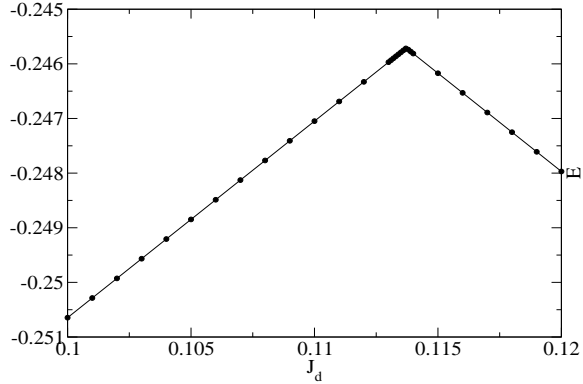}\\
\caption{Ground-state energies for $L=16$ and $J_{z \perp}=0.2$
in the Ising case.}
\label{gsis}
\end{figure}

The curves of $E_G$ in Fig.(~\ref{gs}) show that $E_G$ is differentiable, which 
 suggests that
the transition could be of second order. This is to be constrasted with
the pure Ising equivalent of Hamiltonian (~\ref{hamiltonian}) obtained
by setting all XY terms to $0.$ $E_G$ for the Ising case shown in 
Fig.(~\ref{gsis}) is not differentiable at $J_d^{max}/J_{\perp}$, as
expected since this transition is of first order. It is to be noted that
in this case where quantum fluctuations are absent, $J_d^{max}/J_{\perp}$
does not occur at $0.5$. This is another fact which favors the conclusion
that the deviation from $0.5$ seen in previous studies \cite{sindzingre} 
is a finite size effect rather than some subtle quantum effect due to the 
existence of an intermediate phase between the two N\'eel ordered states.

\subsection{Short-distance spin-spin correlations}


\begin{figure}
\includegraphics[width=3. in, height=2. in]{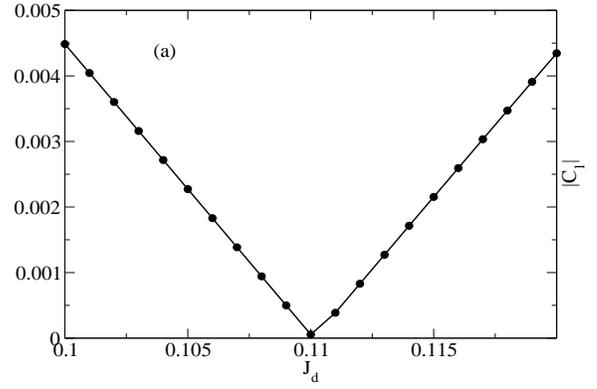}\\
\vspace{1cm}
\includegraphics[width=3. in, height=2. in]{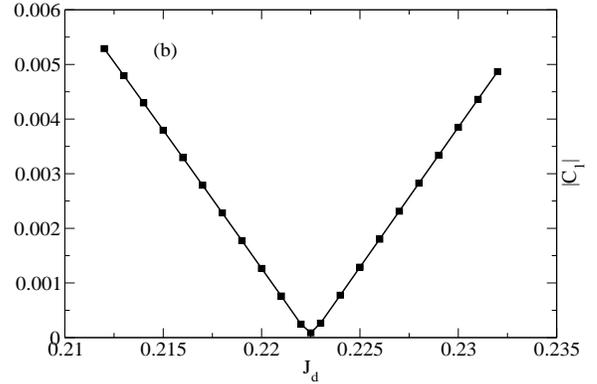}\\
\vspace{1cm}
\includegraphics[width=3. in, height=2. in]{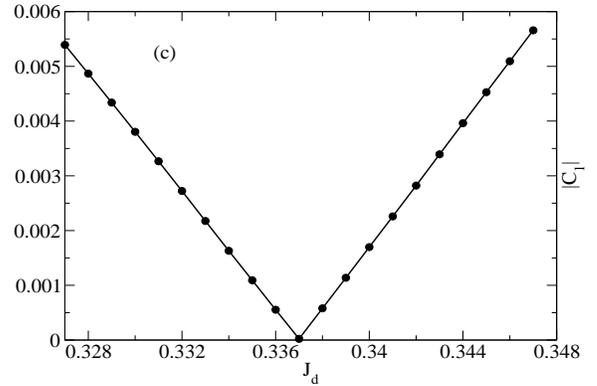}\\
\caption{$C_1$ for $L=16$ and $J_{\perp}=0.2$(a),
$J_{\perp}=0.4$ (b) and, $J_{\perp}=0.6$(c) as function of $J_d$.}
\vspace{0.5cm}
\label{cor1}
\end{figure}

\begin{figure}
\includegraphics[width=3. in, height=2. in]{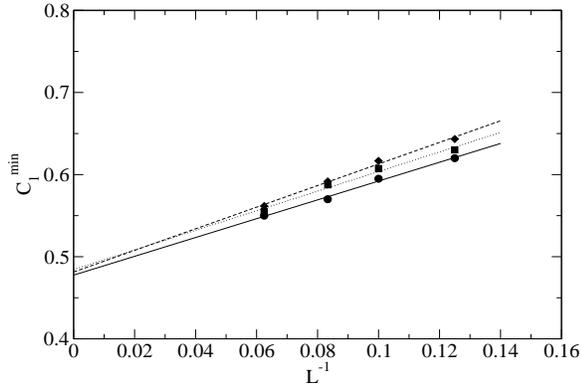}\\
\caption{ Minima of $C_1$ 
 as function of the lattice size for $J_{\perp}=0.2$ (circles),
$J_{\perp}=0.4$ (squares) and, $J_{\perp}=0.6$ (diamonds).}
\vspace{0.5cm}
\label{c1min}
\end{figure}

\begin{table}
\begin{ruledtabular}
\begin{tabular}{ccccc}
 $J_{ \perp}$ & $L=8$ & $L=10$ & $L=12$ & $L=16$  \\
\hline
 $0.2$ & 0.124 & 0.119 & 0.114 & 0.110 \\
 $0.4$ & 0.252 & 0.243 & 0.235 & 0.222 \\
 $0.6$ & 0.386 & 0.372 & 0.359 & 0.337\\
\end{tabular}
\end{ruledtabular}
\caption{ Values of $J_d$ corresponding to the minimum of $C_1$ for 
different $J_{\perp}$ and $L$. }
\label{mincor1}
\end{table}
The first neighbor transverse correlation function (Fig.(~\ref{cor1}))

\begin{equation} 
C_1=\langle S_{il}^zS_{il+1}^z \rangle 
\end{equation}

\noindent taken in the middle
of the lattice $(il)=(L/2,L/2+1)$ is linear in the vicinity  of
$J_d^{max}/J_{\perp}$ and vanishes at the minimum point $J_d^0$. 
Table(~\ref{mincor1}) lists the position of $J_d^0$
for various $J_{\perp}$ and $L$. These values of $J_d^0$ are 
equal to or near $J_d^{max}$ found for $E_G$ for all values of $J_{\perp}$
studied. We believe that the small differences between the values of
$J_d^0$ and $J_d^{max}$ are due to numerical errors on $C_1$ and
$E_G$. The difference between these positions are in all cases less than
$0.1 \%$. Since the DMRG technique tends to have a better accuracy
on the ground-state energies than on the correlation functions,
the position of the maximum in $E_G$  is taken as the reference
for the transition. The position of the minimum shown in 
Fig.(~\ref{c1min}) extrapolate as for $E_G$ towards $J_d/J_{\perp}=0.5$.

Since $C_1=0$ at the transition point, the spins in different
chains remain uncorrelated as in the classical case. This is indeed
the best way to minimize frustration. In order to avoid the cost
in energy induced by frustrated bonds, the system  forms the largest
unfrustrated cluster allowable. In this case, the largest unfrustrated
clusters are independent chains. This state can actually be seen
as the generalization of the Majumdar-Gosh state known in 1D systems
\cite{lhuillier}.
However, this does not mean that
the chains are completely disconnected as in the classical case.
 In the results on $E_G$ above we were unable to find a
point in the parameter space where the ground-state energy of the
coupled 2D system is exactly equal to the ground-state energy of 
disconnected chains. The ground state energies of the 2D systems are 
always lower than that of completely disconnected chains. This means that 
despite the fact that $C_1=0$, all other transverse correlations are  
not necessarily zero at $J_d=J_d^0$ as in the classical case.


\begin{figure}
\includegraphics[width=3. in, height=2. in]{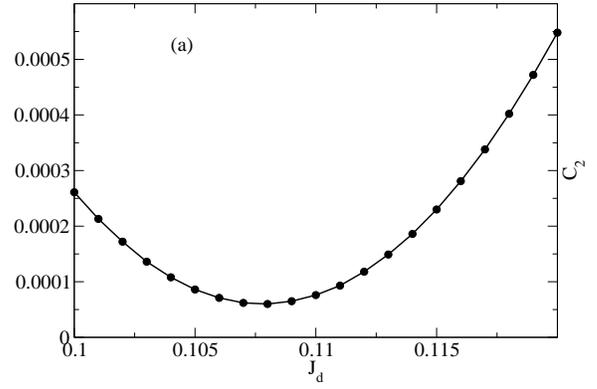}\\
\vspace{1cm}
\includegraphics[width=3. in, height=2. in]{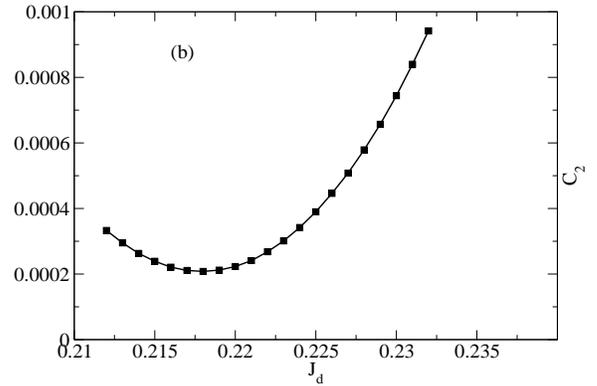}\\
\vspace{1cm}
\includegraphics[width=3. in, height=2. in]{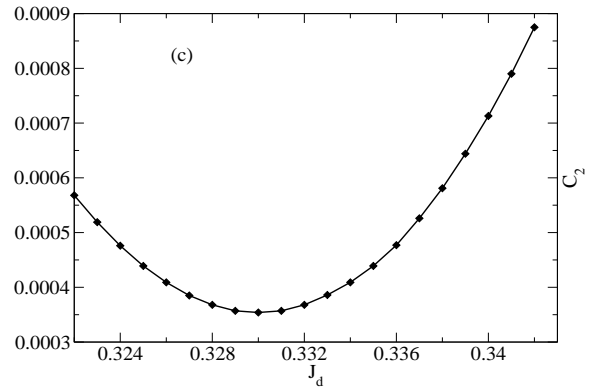}\\
\caption{$C_2$ for $L=16$ and $J_{\perp}=0.2$(a),
$J_{\perp}=0.4$ (b) and, $J_{\perp}=0.6$ (c) as function of $J_d$.}
\vspace{0.5cm}
\label{cor2}
\end{figure}

\begin{figure}
\includegraphics[width=3. in, height=2. in]{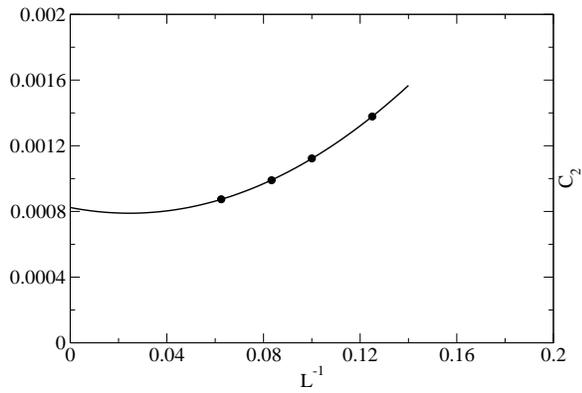}\\
\caption{ Value  of $C_2$ at the minimum
 as function of the lattice size for $J_{\perp}=0.6$ (circles).}
\vspace{0.5cm}
\label{c2min}
\end{figure}

The second neighbor correlation
\begin{equation}
 C_2=\langle S_{il}^zS_{il+2}^z \rangle
\end{equation}

\noindent shown in Fig.(~\ref{cor2}),
taken at $(il)=(L/2,L/2+2)$, does not vanish. It instead has a minimum
at $J_d^{min}$ for all values of $J_{\perp}$. The values of $C_2$ 
corresponding to $J_d=J_d^0$ shown in Fig.(~\ref{c2min}) for $J_{\perp}=0.6$
extrapolate to a finite value in the thermodynamic limit. Thus, at the
point where $C_1=0$, $C_2$ has a small finite value. It is worth noting
that the dimerization is not the only way to avoid frustration as it is
commonly believed \cite{lhuillier}.

\subsection{Dimerization}

\begin{figure}
\includegraphics[width=3. in, height=2. in]{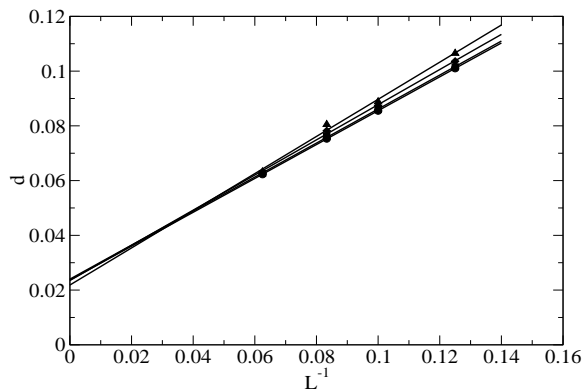}\\
\caption{Evolution of the dimerization as function of the lattice size
for  $J_{\perp}=0.$ (circles), $J_{\perp}=0.2$ (squares),
$J_{\perp}=0.4$ (diamonds) and, $J_{\perp}=0.6$ (triangles).}
\label{dim}
\end{figure}

The examination of the short-range correlations above proves that at
the maximally frustrated point, there is a weak bond between the chains.
The ground-state in this region of parameters has been predicted via 
field theoretic approaches \cite{tsvelik,starykh} to be a columnar dimer
state, with the dimerization occuring along the chains. 
In order to see if dimerization occurs along the chains, we
compute the quantity

\begin{equation}
d_{il}=\langle S_{2il}^zS_{2i+1l}^z \rangle-\langle S_{2i+1l}^zS_{2i+2l}^z
 \rangle.
\end{equation}

\noindent One should note that since we use the OBC, we are in the most
favorable situation for the dimerization. i.e., the OBC breaks the translational
symmetry of the chains and even isolated finite chains are dimerized. This
spurious dimerization decays very slowly as $L \rightarrow \infty$. It is
thus expected that if indeed the system dimerizes at the QCP, the finite
size behavior of $d_{il}$ of the coupled system will be different from that
of isolated chains. In Fig.(~\ref{dim}) $d$ is shown for $J_{\perp}=0$,
$0.2$, $0.4$, and $0.6$ and for various sizes at $J_d^{max}$ in the relevant
cases. $d$ converges to the same limit whether $J_{\perp}=0$ or not. Since
the single chain is not dimerized, this shows that the 2D system at the
critical point is not dimerized either.

We are aware of a different conclusion reached by a recent bosonization
approach coupled to a renormalization group and mean-field 
analyses \cite{starykh}. In that study for a two-leg ladder in the
vicinity of the maximally frustrated point, it was found that 
 although the action of $J_{\perp}$ is largely cancelled by $J_d$, 
at the second order there remains residual couplings 
${\tilde J}_1=2J_d^2/\pi^2$ and 
${\tilde J}_2=-3J_d^2/\pi^2$ which are relevant. ${\tilde J}_1$ favors 
N\'eel order and ${\tilde J}_2$, the ring-exchange term, favors a dimerized 
state. These authors then carried out a self-consistent mean-field 
approximation of the competition between ${\tilde J}_1$ and ${\tilde J}_2$
in the 2D system. They found a narrow dimerized state between the two
ordered magnetic states. According to their conclusions, our prediction
of the spin liquid state at the transition is due to finite size effects.
But our results above, performed for larger interchain couplings for 
which the dimerization should be more apparent and is not seen, rather
support our conclusions in Ref.~\cite{moukouri-TSDMRG2}. There is  
additional evidence, discussed below, obtained from the study of the
three-leg ladder\cite{wang} which also agrees with the absence a dimer phase.
It is important mentioning that this type of RG analysis is not without 
a certain bias. The RG transformation usually leads to a proliferation of 
couplings. There is often some degree of arbitrariness in the choice of the
relevant couplings.

\subsection{Long-distance spin-spin correlations}

\begin{figure}
\includegraphics[width=3in, height=2. in]{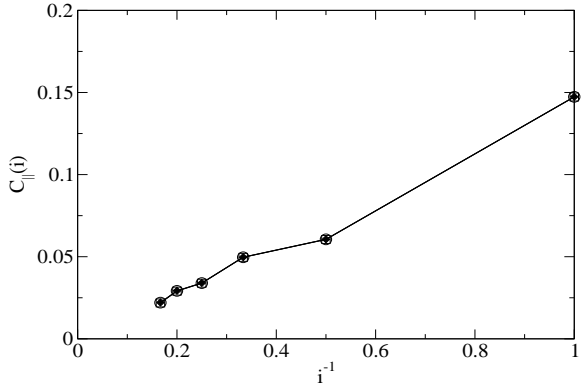}
\caption{Longitudinal spin-spin correlation at $J_d^{max}$ as function 
of the distance for $L=16$ and, $m=32$ (circles), 
$m=64$ (squares) and, $m=96$ (diamonds).} 
\vspace{0.5cm}
\label{corlg0.2par}
\end{figure}

\begin{figure}
\includegraphics[width=3in, height=2. in]{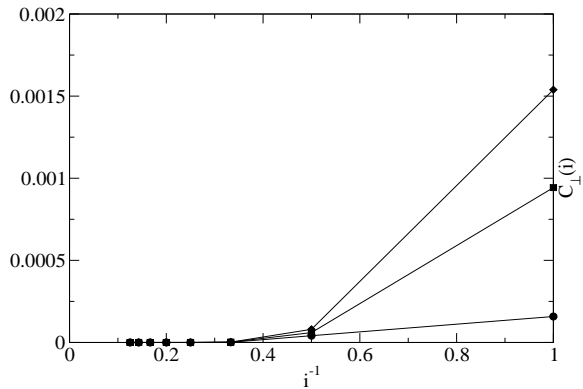}
\caption{Transverse spin-spin correlation at $J_d^{max}$ as function of the distance 
for $L=16$ and, $m=32$ (circles), $m=64$ (squares) and, $m=96$ (diamonds).} 
\vspace{0.5cm}
\label{corlg0.2tran}
\end{figure}

Near the maximally frustrated point, the long distance behavior of
the in-chain spin-spin correlation functions is nearly identical
to that of the disconnected chains. These correlations shown in
Fig.(~\ref{corlg0.2par}) for $J_{\perp}=0.2$ and $L=16$ are nearly
independent of $m$. This is because, for a purely 1D system of this 
size they are already known with very good accuracy. Since
the chains are nearly disconnected, no significant increase of $m$ is
required to describe them with the same level of accuracy as in 1D.

This conclusion is supported by the behavior of the transverse 
correlation functions shown in Fig.(~\ref{corlg0.2tran}) for the same
set of parameters. $C_{\perp}(i)$ is found to be equal to $0$ for
$i \agt 4$ for all the values of $m$.  

\begin{figure}
\includegraphics[width=3. in, height=2. in]{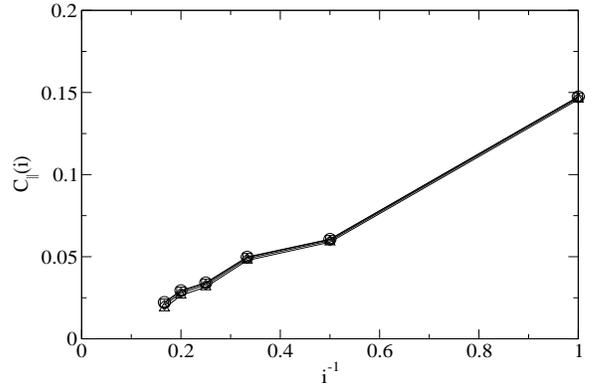}
\caption{Longitudinal spin-spin correlation at $J_d^{max}$ 
as function of the distance
for  $J_{\perp}=0.$ (circles), $J_{\perp}=0.2$ (squares),
$J_{\perp}=0.4$ (diamonds) and, $J_{\perp}=0.6$ (triangles).}
\vspace{0.5cm}
\label{corlm96par}
\end{figure}

\begin{figure}
\includegraphics[width=3. in, height=2. in]{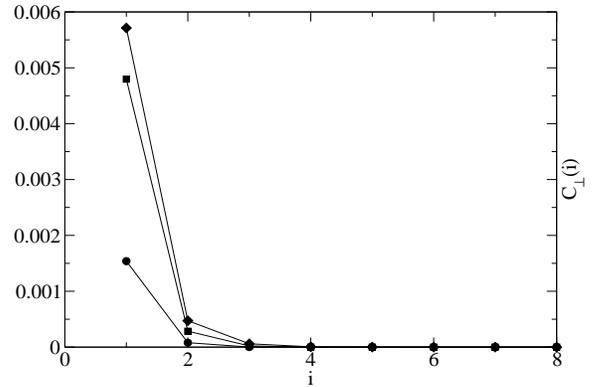}
\caption{Transverse spin-spin correlation at $J_d^{max}$
as function of the distance for  $J_{\perp}=0.2$ (circles), $J_{\perp}=0.4$ 
(squares) and, $J_{\perp}=0.6$ (diamonds).} 
\vspace{0.5cm}
\label{corlm96tran}
\end{figure}

 $C_{\parallel}(i)$ and $C_{\perp}(i)$ are independent of
$J_{\perp}$ at the maximally frustrated point, i.e. only the ratio
$J_d/J_{\perp}$ is important up to intermediate values. This is seen 
in Fig.(~\ref{corlm96par})
and Fig.(~\ref{corlm96tran}) where these quantities are compared for 
$J_{\perp}=0.2$, $0.4$ and $0.6$. $C_{\parallel}(i)$ is nearly identical
to that of independent chains while $C_{\perp}(i)$ remains very small and
decays exponentially with $i$. This behavior is valid from weak to 
intermediate coupling regimes. It is to be noted that since the maximally
frustrated point is chosen at the maximum $J_d^{max}$ of $E_G$, $C_1$ is 
not equal to $0$ because the two quantities differ slightly as shown in
Table(~\ref{maxgs},~\ref{mincor1}) because of numerical uncertainties.  

\begin{figure}
\includegraphics[width=3. in, height=2. in]{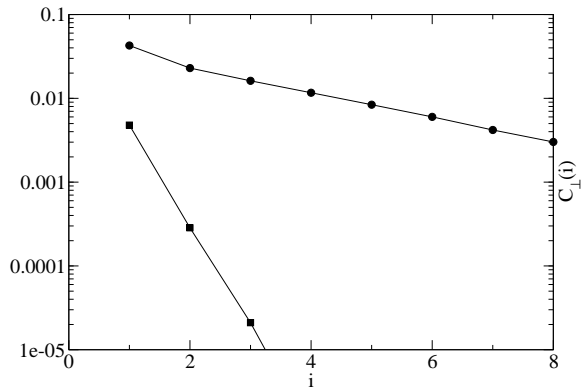}
\caption{Transverse spin-spin correlation as function of the distance
for  $J_{\perp}=0.4$, $J_d=0$ (circles), $J_{\perp}=0.4$, $J_d=0.219$ (squares).
}
\vspace{0.5cm}
\label{corlm96g04tran}
\end{figure}

 This behavior of $C_{\perp}(i)$ at
$J_d \approx J_{\perp}/2$ is to be contrasted to that of the  magnetic regime.
In  Fig.(~\ref{corlm96g04tran}) we see that for
instance, $C_{\perp}(4)$ of the magnetic case is already four orders of
magnitude larger than in the disordered case. The exponential decay 
of $C_{\perp}(i)$ seen in
Fig.(~\ref{corlm96g04tran}) and the decay of $C_{\parallel}(i)$ which is nearly
that of decoupled chains suggest that for $J_d \approx J_{\perp}/2$,
the system displays a spin analog of the sliding Luttinger liquid (SLL)
discussed in fermionic models \cite{carpentier, kivelson, kane}. In the
region $J_d \approx J_{\perp}/2$ the two  competing magnetic fluctuations $(\pi,\pi)$
and $(\pi,0)$ cancel each other leading to irrelevant interchain couplings.
These irrelevant couplings renormalize the energy but do not modify
the behavior of the correlation functions. Our findings show unambiguously
how the SLL concept is linked to quantum criticality. It
also provides evidence of fractionalization at the critical point as
first suggested in Ref\cite{laughlin}.

\subsection{Finite size spin gap}

\begin{figure}
\includegraphics[width=3. in, height=2. in]{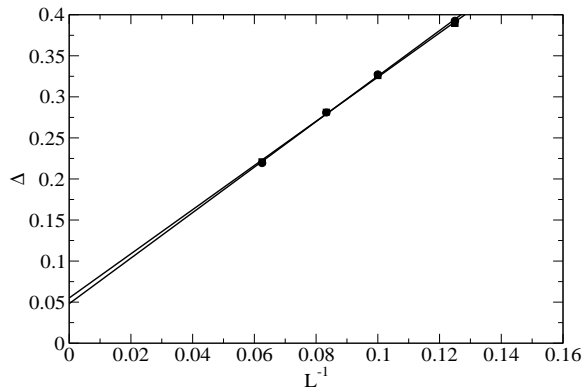}
\caption{Finite size gap with OBC as function of the lattice
size for $J_{\perp}=0$ (circles) and, $J_{\perp}=0.2$ (squares).}
\label{gap}
\end{figure}

 Logarithmic corrections and the spurious dimerization introduced by the 
OBC is accompanied by
a spurious gap in the pure 1D system. In Fig.(~\ref{gap}) we see that
at the maximally frustrated point, the gap for $J_{\perp}=0.2$ converges
 to the same value as that of $J_{\perp}=0$. It is thus reasonable,
knowing that the single chain is gapless, to conclude that the 2D
system is also gapless in the thermodynamic limit. We note that the
logarithmic corrections are not taken into account in the gaps predicted
by exact diagonalization studies\cite{sindzingre}.

\subsection{Implications for the isotropic $J_1-J_2$ model}

For the isotropic case ($J_{\perp}=J$) it is usually said that
there is a large amount of evidence supporting the fact that
the magnetic order vanishes for $0.4 \alt J_d \alt 0.6$, i.e., a
spin gapped state exists in this region \cite{lhuillier}. But in
our opinion, this conclusion is not that compelling. The results
displayed above indicate that the limit $J_d/J_{\perp} \rightarrow 0.5$
is in fact analogous to the problem of coupled unfrustrated chains ($J_d=0$ in
Hamiltonian(~\ref{hamiltonian})) in the limit $J_{\perp} \rightarrow 0$. 
For this latter problem there was a prediction of a spin liquid state for
$J_{\perp} \alt 0.1$ from ED, field theory mapping and spin wave approaches  
\cite{parola}. However, a QMC study \cite{sandvik} later showed that the
N\'eel order extends down to $J_{\perp}=0.01$. It is now believed that
long-range order sets in as soon as $J_{\perp} \neq 0$. This suggests that
the disordered phase  often claimed to exists for the isotropic $J_1-J_2$
model, which was obtained from ED of only a maximum of 
$6 \times 6$ systems or series expansions such as dimers expansions which 
may be biased toward dimerization, is not a settled  issue.  

In fact there is a strong evidence, unfortunately neglected, which points to
the contrary. A DMRG study \cite{wang} performed on a three-leg ladder 
found that the system is gapless for all values of $J_{\perp}$ and $J_d$.
The lenght of the chains were up to $L=360$ so that finite size effects
were negligible.  The computed phase diagram has two states a symmetric
doublet phase and the quartet phase. The low energy properties of the
symmetric doublet phase are similar to those of a single spin-$1/2$ chain.
The scaling behavior of the quartet phase is identical to that of a
spin-$3/2$. These two states will naturally evolve towards the N\'eel
$Q=(\pi,\pi)$ and $Q=(\pi,0)$, respectively, when the number of chains
goes to infinity. This is  fully consistent with our conclusions. 

The perturbative method applied in this study cannot be used directly to
study the isotropic case, since  starting with the decoupled chains is too
biased. But if the principles uncovered in the
anisotropic case are to be applied, it appears that even in
the isotropic situation, the nearly decoupled chains ground state
is still the best way to minimize frustration. From this consideration,
we applied the method to the isotropic case and arrived at the same
conclusions as above. Namely that the ground state at the maximally 
frustrated point is made of nearly disconnected chains and is thus 
a SLL.

Another  indication for this is the finding  by a recent ED 
study \cite{sindzingre} that even at $J_{\perp}=0.8J$ the behavior
observed at lower coupling still persists. However, this study also
predicts a spin gap. We believe  this apparent contradiction is
due to the fact that the ED study was done with an even number of chains
and the logarithmic corrections were not considered.

\section{Doped systems}

Having identified the ground state at half-filling as a spin version
of a SLL, we now turn to the doped case. We will mostly restrict ourselves
 to the
vicinity of the maximally frustrated point $J_d \approx J_{\perp}/2$.
This is primarily because the results at half-filling indicate that
far from this point no significantly new physics is likely to occur. One
would expect that the two N\'eel ground states found at half-filling
away from the the QCP will naturally evolve into spin density wave (SDW)
ground states upon doping. The really physically interesting question
is: What is the effect of the doping upon the spin SLL?

The main finding in the doped case is that the magnetic properties of the
system do not significantly change as long as one remains close to 
half-filling. But if the hole doping becomes too large, $x \approx 0.3$,
the model displays a qualitative change. The frustration ceases, and $J_d$
cooperates instead of competing with $J_{\perp}$.

\subsection{Ground-state energies}

\begin{figure}
\includegraphics[width=3. in, height=2. in]{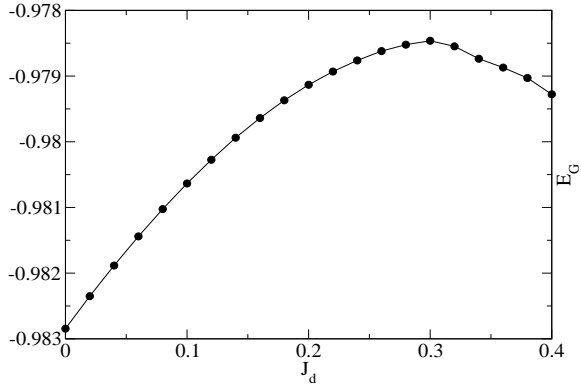}
\caption{Ground state energy for $x=0.167$, $L=12$, and $J_{\perp}=0.4$ as 
function of  $J_d$.}
\vspace{0.5cm}
\label{egd0.833}
\end{figure}

\begin{figure}
\includegraphics[width=3. in, height=2. in]{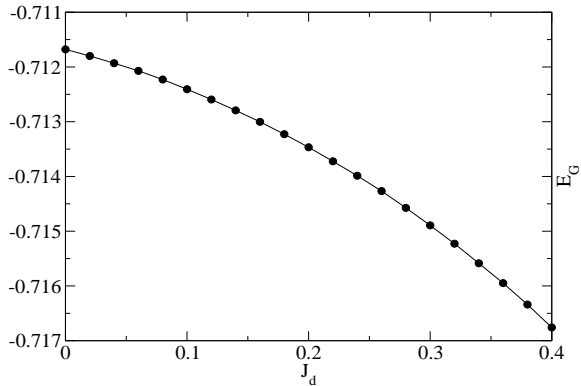}
\caption{Ground-state energy for $x=0.4$, $L=10$, and $J_{\perp}=0.2$ 
as function of  $J_d$}
\vspace{0.5cm}
\label{egd0.6}
\end{figure}

$E_G$ at $x=0.167$ for $L=12$ as function of $J_d$ is
shown in Fig.(~\ref{egd0.833}). It displays a maximum as for the half-filled
case. One can note that this maximum is now shifted towards
higher values of $J_d$. This maximum goes from $J_d^{max}=0.236$ for $x=0$ to
$J_d^{max}=0.30$ for $x=0.167$. By inserting the holes, the system becomes
less frustrated. It is therefore necessary to increase $J_d$ in order to balance the
action of $J_{\perp}$. We find that if the hole density is increased above
$x \approx 0.3$ the maximum in $E_G$ is suppressed. For $x \agt 0.3$, $E_G$
always decreases with $J_d$. This is seen for instance in Fig.(~\ref{egd0.6})
where $E_G$ for $x=0.4$ is shown. This shows that for large dopings,
contrary to the near half-filling situation, 
$J_d$ increases instead of reducing the stability of the system.

\subsection{Short-distance spin-spin correlations}

\begin{figure}
\includegraphics[width=3in, height=2.in]{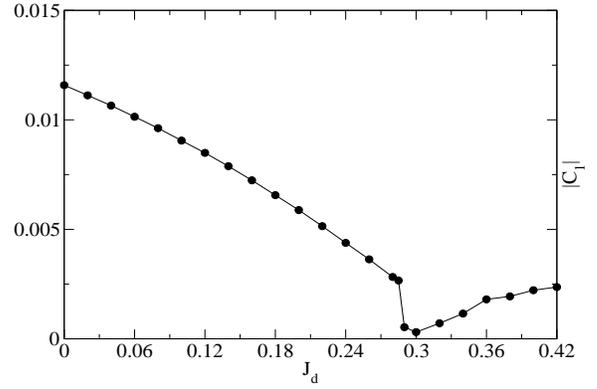}
\caption{$C_1$ for $x=0.167$, $L=12$, and $J_{\perp}=0.4$ as
function of  $J_d$}
\vspace{0.5cm}
\label{cor1d0.833}
\end{figure}
                                                                                
\begin{figure}
\includegraphics[width=3. in, height=2. in]{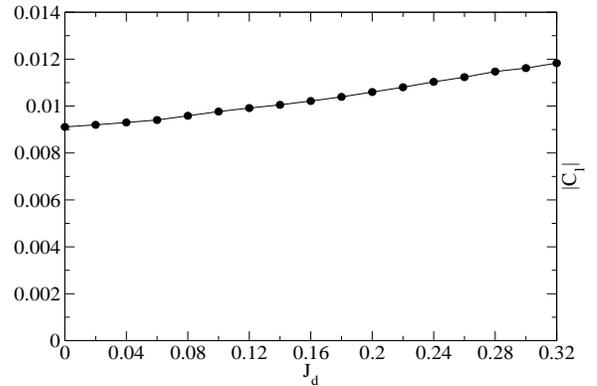}
\caption{$C_1$ for $x=0.4$, $L=10$, and $J_{\perp}=0.2$
as function of  $J_d$}
\label{cor1d0.6}
\vspace{0.5cm}
\end{figure}

This difference between the near half-filling and larger hole densities
is also displayed in $C_1$. In Fig.(~\ref{cor1d0.833}) corresponding to
$x=0.167$, $C_1$ goes to $0$ at $J_d^0 =0.3$. One may note that there
seems to be a jump at the transition point. This jump was not observed
at half-filling. This could indicate a possible first order transition.
However this is not supported by the behavior of $E_G$ shown above. We
notice during the simulations that, if one is too close to the transition
point, for the same set of parameters, one can go from one side of the 
transition when the number of chains is small to the other side when the 
number of chains gets larger.  We thus believe that the jump observed in 
$C_1$ is somehow related to the reduced accuracy in the doped case induced
by the mixture of the two competing ground states near the transition
point. The regularity of the curve far from the transition point justify
this assumption. In agreement with the behavior of $E_G$, the
transition is suppressed when the hole doping is too large. $C_1$ shown
in Fig.(~\ref{cor1d0.6}) always remain antiferromagnetic and increases
with $J_d$.   

\subsection{Long-distance correlations}
                               
\begin{figure}
\includegraphics[width=3. in, height=2. in]{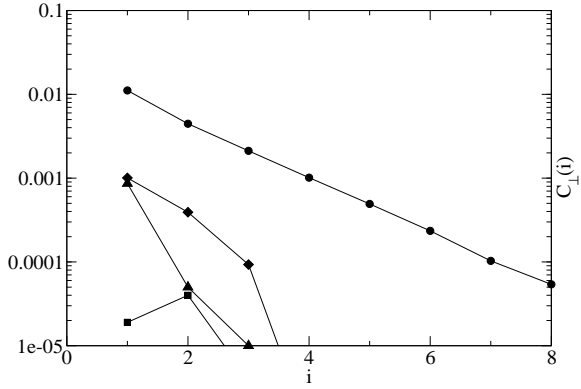}
\caption{Transverse spin-spin correlation as function of the distance
for  $J_{\perp}=0.2$, $J_d=0$ (circles), and at the maximally frustrated
point at $J_{\perp}=0.2$ (squares), $J_{\perp}=0.4$ (triangles), 
$J_{\perp}=0.6$ (diamonds).}
\vspace{0.5cm}
\label{corld0.875l16tran}
\end{figure}

$C_{\perp}(i)$ shown in Fig.(~\ref{corld0.875l16tran}) for $x=0.125$ 
retains essentially the
same behavior as in the half-filled case. It decays exponentially at the 
maximally frustrated point for $J_{\perp}=0.2$, $0.4$, and $0.6$ while the
decay is significantly slower in the unfrustrated case. In the absence of 
frustration, $C_{\perp}(i)$ has an antiferromagnetic signature while 
$C_{\parallel}(i)$ is SDW like with 
$k_F=\frac{\pi}{2} (1-x)$, where $x$ is the
density of holes. Though we have not made a finite size analysis, it is 
reasonable to believe that the system is in an SDW ground state in the 
thermodynamic limit. At the maximally frustrated point, the magnetic
order parameter vanishes.

\begin{figure}
\includegraphics[width=3. in, height=2. in]{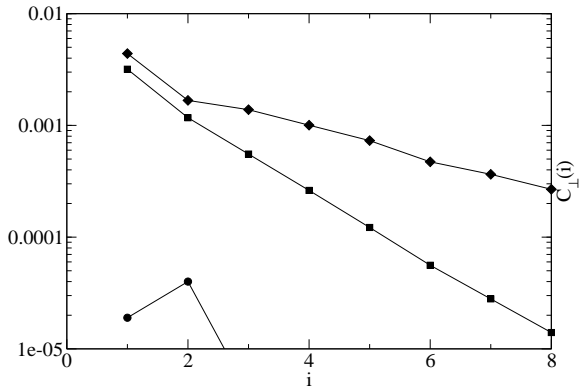}
\caption{Transverse spin-spin correlation as function of the distance
for  $J_{\perp}=0.2$,$J_d=0.11$ for $x=0.125$ (circles), 
$x=0.25$ (squares) and $x=0.375$ for $L=16$.}
\label{gld16tran}
\end{figure}

\begin{figure}
\includegraphics[width=3. in, height=1. in]{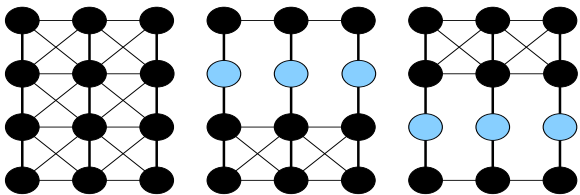}
\caption{Holes (light color) configurations that minimize frustration 
for $x=0.25$. The left illustration represents the half-filled case where 
the system is fully frustrated. }
\label{frvsholes}
\end{figure}

 As $x$ is further increased to $0.167$, $0.25$, and  $0.375$, we find that 
 the role of frustration is reduced. Ultimately, the
system evolves towards a conventional SDW ground state. 
This is seen in Fig.(~\ref{gld16tran}) where the exponential decay of 
$C_{\perp}(i)$  is significantly slowed at large hole dopings.
There is already four orders of magnitudes between $C_{\perp}(4)$ for
$x=0.125$ and $x=0.375$. It  appears that the system tends to choose hole 
configurations that minimize frustration in order to avoid
 the cost in energy induced by frustration. It is in the vicinity of $x=0.125$ 
that frustration is most effective, i.e., the magnetic order which exists in 
the non-frustrated case can easily be cancelled by frustration. Beyond this 
point, frustration becomes less and less effective. At quarter filling the 
system can choose a hole configuration so that frustration becomes  
ineffective. Such possible configurations are shown in Fig.~(\ref{frvsholes}), 
but we have not made a systematic analysis of them. Since the introduction
of holes usually tends to reduce the order parameter, the reemergence of
magnetism at the maximally frustrated point by introduction of holes
can be viewed as the manifestation of Villain's order from disorder effect
\cite{villain}.

\begin{figure}
\includegraphics[width=3. in, height=2. in]{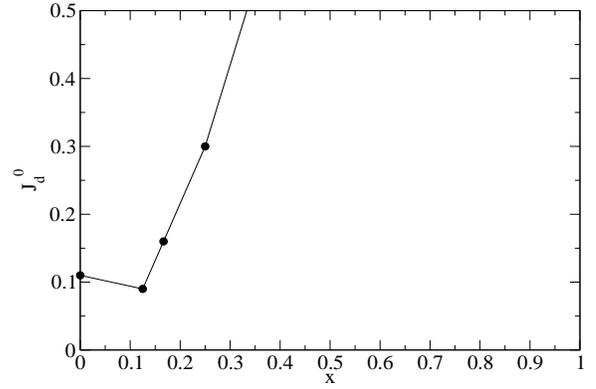}
\caption{Critical values of $J_d$ as function of doping.}
\label{critdop}
\end{figure}

These arguments are illustrated in Fig.(~\ref{critdop}) where the value
of $J_d^0$ is shown as function of $x$. $J_d^0(x)$ first slowly decreases
from half-filling until about $x=0.125$ and then sharply rises. Thus
there is a point, for $x$ between $0.25$ and $0.5$, where the critical
point is suppressed.

\subsection{Equal-time Green's function}

The transition point is also signaled in the equal-time transverse Green's
function

\begin{eqnarray}
G_{\perp}(i)= \langle c_{jl\sigma}c^{\dagger}_{jl+i \sigma} \rangle.
\end{eqnarray}

\begin{figure}
\includegraphics[width=3. in, height=2. in]{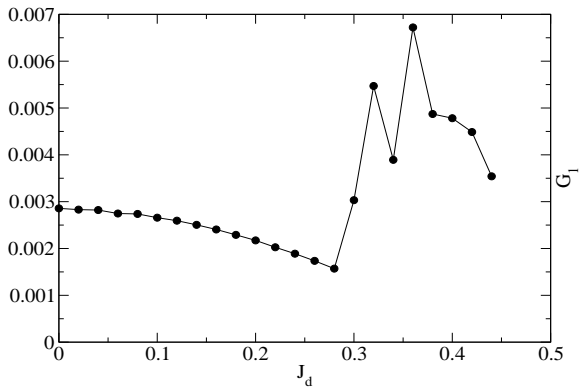}
\caption{Transverse equal-time first neighbors Green's  function at 
$x=0.167$ as function of  $J_d$ for $L=12$.}
\vspace{0.5cm}
\label{g1f16tran}
\end{figure}

$G_1=G_{\perp}(1)$ is shown in Fig.(~\ref{g1f16tran}) decays from $J_d=0$ to $J_d=J_d^{max}$. This
decay of $G_1$ is consistent with the tendency to localize the 
particles within the chains induced by $J_d$. However, in the
vicinity of $J_d^{max}$, the situation becomes less clear. $G_1$
strongly oscillates. These oscillations cease when $J_d$ is beyond
the critical region where $G_1$ decays again. It appears that
$G_1$ behaves differently in the two  ordered states; it increases from the
transition point in the first and decreases from the transition point in the
second state. The curious behavior seen in the transition region could
well be the manifestation of the mixture of the two ground states 
seen in $E_G$ and in $C_1$. This mixture affects the value of
fermion correlation functions more severely than their spin counterpart.
For this reason, we have not calculated  charge density or superconducting 
correlations which involve four fermions. 
There is however no reason to suspect a charge density or superconducting
ordering in the vicinity of $J_d^{max}$. These transition would be of first
order and would thus be apparent in the ground state energy. 

\begin{figure}
\includegraphics[width=3. in, height=2. in]{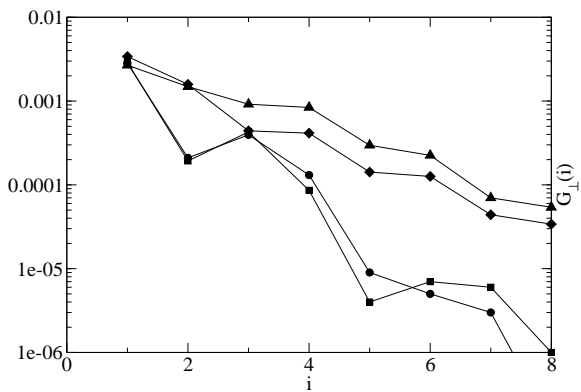}
\caption{Transverse equal-time Green's  function at $x=0.125$
for  $J_d=0.2$, $J_d=0$ for spin up (circles), spins down (squares);
$J_{\perp}=0.2$,$J_d=0.11$ for spins up (triangles), spins up (diamonds)}
\label{glf16tran}
\end{figure}

Despite the uncertainties caused by the mixture of the two ground states on
the two sides of the transition, we nevertheless show in Fig.(~\ref{glf16tran}) 
the equal-time transverse Green's function in the SDW case and in the 
magnetically disordered case. The decay in the disordered state is slower 
than in the magnetic case. This raises the possibility of the relevance of
the single particle hopping. If true, this would mean that the spin and the
charge behave differently in the transverse direction at the transition
point. More work needs to be done in order to clarify this point. 

\section{Conclusions} 

In this study, we have presented numerical evidence for the connection 
between quantum phase transition and Luttinger liquid physics from the study
of coupled $t-J$ chains.  The ordered $Q=((1-x)\pi,\pi)$ N\'eel state that 
exists at half-filling and near half-filling for $J_{\perp} \neq 0$ and $J_d=0$ 
is destroyed at $J_d=J_d^{max}$. At this point, the in-chain spin-spin 
correlation of the 2D system decay nearly like those of independent chains. 
A careful examination of the transverse spin-spin correlations reveals that 
the chains are loosely bound and that these correlations decay exponentially. 
We thus identify the state of the system at this maximally frustrated point 
as a sliding Luttinger liquid. These properties are observed even when the 
interchain coupling are well beyond the perturbative regime

The key mechanism behind this connection is magnetic frustration. The 
leading effect which brings a compromise between the frustrated bonds
is the severing of these bonds to form the largest unfrustrated clusters
with the lowest energy. These unfrustrated units are then coupled by some 
residual interactions generated by quantum fluctuations. These residual 
interactions may or may not drive the system away from the physics displayed 
by the largest unfrustrated clusters. This picture is in constrast to the 
widely accepted hypothesis which suggests that the leading mechanism to 
avoid frustration is short or long-range dimerization which may (spin-Peierls) 
or may not (Resonating Valence Bond) lead to the breaking of the lattice 
translational symmetry. It is however to be noted that the two pictures do 
not necessarily always conflict with each other. In some models, the largest 
unfrustrated clusters could just be dimers or plaquettes. Our point here is 
that this is not the case in the anisotropic $J_1-J_2$ model.

In the model studied in this paper, LL physics arises at the QCP because the 
severing of frustrated bonds leads to nearly independent chains which are
the largest unfrustrated clusters. These chains are coupled, even when the 
original interchain couplings are large, only by  residual
interactions which do not appear to destabilize the 1D physics at the 
QCP. In a numerical simulation, it is however impossible to rule out 
completely the emergence of relevant interchain interactions at a very 
low energies.

\begin{acknowledgments}
 We wish to thank M.E. Fisher, C.L. Kane, S. Kivelson, R.B. Laughlin, and
 A.-M.S. Tremblay for helpful discussions.  We thank S. Haas for sharing
his $t-J$ ladder data. We also thank J. Allen for numerous exchanges during 
the course of this work. We are grateful to P.L. McRobbie for reading the
manuscipt.
\end{acknowledgments}

\end{document}